\documentclass[reprint,
showpacs,preprintnumbers,
 amsmath,amssymb,aps,
 prl,
]{revtex4-1}

\usepackage{subfigure}
\usepackage[rflt]{floatflt}
\usepackage{color}
\usepackage{xspace}

\usepackage{graphicx}
\usepackage{dcolumn}
\usepackage{bm}

\definecolor{darkblue}{rgb}{0.149019,0,0.439215}
\usepackage[dvipdfm,colorlinks=true,citecolor=darkblue,linkcolor=darkblue,urlcolor=darkblue]{hyperref}

\newcommand{\lin}{PbCuSO$_4$(OH)$_2$\xspace}

\begin{document}


\title{Complex field-induced states in Linarite \lin with a variety of high-order exotic SDW$_p$ states}

\author{B. Willenberg$^{1,2}$, M.\ Sch\"{a}pers$^{3}$, A.U.B.\ Wolter$^3$, S.-L.\ Drechsler$^3$, M.\ Reehuis$^2$, B.\ B\"{u}chner$^{3,4}$, A.J.\ Studer$^5$, K.C.\ Rule$^5$, B.\ Ouladdiaf$^6$, S.\ S\"{u}llow$^1$, and S.\ Nishimoto$^{3,4}$}

\address{
$^1$Institute for Condensed Matter Physics, TU Braunschweig, D-38106 Braunschweig, Germany\\
$^2$Helmholtz Center Berlin for Materials and Energy, D-14109 Berlin, Germany\\
$^3$Leibniz Institute for Solid State and Materials Research IFW Dresden, D-01171 Dresden, Germany\\
$^4$Institut f\"ur Festk\"orperphysik, TU Dresden, D-01062 Dresden, Germany\\
$^5$The Bragg Institute, ANSTO, Kirrawee DC NSW 2234, Australia\\
$^6$Institute Laue-Langevin, F-38042 Grenoble Cedex, France}

\date{\today}

\begin{abstract}

Low-temperature neutron diffraction and NMR studies of field-induced
phases in linarite are presented for magnetic fields $H \parallel b$
axis. A two-step spin-flop transition is observed as well as a
transition transforming a helical magnetic ground state into an
unusual magnetic phase with sine-wave modulated moments $\parallel
H$. An effective $\tilde{J}_1$-$\tilde{J}_2$ single-chain model with
a magnetization-dependent frustration ratio $\alpha_{\rm eff} =
-\tilde{J}_2/\tilde{J}_1$ is proposed. The latter is governed by
skew interchain couplings and shifted to the vicinity of the
ferromagnetic critical point. It explains qualitatively the
observation of a rich variety of exotic (for strongly correlated
cuprate spin-1/2 Heisenberg systems) longitudinal collinear
spin-density wave SDW$_p$  states ($9 \geq p \geq 2$).
\end{abstract}

\pacs{75.10.Jm, 75.25.-j, 75.30Kz, 75.40Cx}

\maketitle

Recently, frustrated spin chains with ferromagnetic nearest
neighbor $J_1$ (FM-NN) and antiferromagnetic 2$^{\rm nd}$ neighbor
$J_2$ (AFM-NNN) exchange have been discussed in the context of
novel states of matter. Close to the saturation field by tuning
the frustration ratio $\alpha = -J_2/J_1$ of an isotropic model, a
sequence of distinct spin-multipolar (MP) phases should develop
with well-defined phase boundaries, which are described as a
Tomonaga-Luttinger-liquid of $p$-magnon bound states
\cite{Kecke2007,Hikihara2008,Sudan2009,Heidrich-Meisner2009,Zhitomirsky2010,Nishimoto2010,Nishimoto2013,Sato2013,starykh2014}.
They compete with exotic longitudinal spin density wave (SDW$_p$)
correlations, which should prevail in lower magnetic fields.
Interchain exchange should weaken the MP correlations, while
magnetic anisotropy should stabilize them
\cite{Nishimoto2010,Nishimoto2013,Sato2013,starykh2014}.

A proof of existence for spin-MP ordering in real quasi-1D materials
is still lacking. The FM-AFM chain $J_1$-$J_2$ compound LiCuVO$_4$
was considered as a candidate, undergoing a transition into an
incommensurate (ICM) helical phase below 2.1\,K, into a spin-flop
phase in fields of $\sim$\,2.5\,T and an exotic SDW$_2$ phase above
$\sim$\,8\,T \cite{Gibson2004,Buttgen2007,Buttgen2010,Mourigal2012}.
A shift of the SDW$_2$ propagation vector, consistent with
longitudinal density waves of bound $p=2$-magnons, was reported for
low fields of 8--14.5\,T, together with a transition from long- to
short-range magnetic order \cite{Mourigal2012}. It was interpreted
as a signature of coexisting SDW$_2$ and bond-nematic order, a view
disputed in Refs.~\cite{starykh2014,starykh2015} and instead related
to a pinned SDW$_2$. Via magnetization and NMR it was concluded that
MP correlations in LiCuVO$_4$ can exist only in a narrow high field
range $\sim 40$\,T \cite{Zhitomirsky2010,Svistov2011,Buttgen2014}.

The issue not resolved in this dispute is the relationship between
SDW and MP in the long-range ordered phases appearing in 2D and 3D
(in 1D the precursor "phases" overlap \cite{Sudan2009}).
Theoretically, the possibility of homogeneously coexisting SDW$_2$
and nematic phase and/or phase separation has been suggested for
the isotropic model in an extreme quasi-1D regime for specific
intrachain and very weak 2D FM interchain couplings based on
perturbative scattering theory \cite{Ueda2014}. In contrast, only
a 1$^{\rm st}$-order phase transition was predicted for the same
model \cite{starykh2014,starykh2015}. Also predicted is a phase
separation and/or a 1$^{\rm st}$-order phase transition between
nematic and FM phase in a 3D $bcc$ structure for the same
$J_1$-$J_2$ model and a similar approach \cite{Ueda2013} as in
Ref.~\cite{Ueda2014}. The situation is far from clear in
LiCuVO$_4$. The phase diagram has not been studied in detail due
to the high fields required to access it \cite{Buttgen2012}.
Further, the influence of defects on the magnetic properties is
not well understood
\cite{Prozorova2015,Buttgen2014,Prokofiev2004}. Thus, what is
lacking in this context is a comprehensive study of a clean
frustrated FM-AFM spin chain material to properly define these
issues.

A unique example of a frustrated spin chain system for such studies
is linarite. It crystallizes in the monoclinic space group $P2_1/m$
\cite{Effenberger1987}, forming buckled CuO$_2$ chains along the $b$
axis. These have been modelled as a $s = \frac{1}{2}$ spin-chain
with FM-NN $J_1= -100$\,K and AFM-NNN $J_2= 36$\,K
\cite{Wolter2012}. In this $J$-parameter range the saturation field
is about 10\,T, allowing full experimental access to the magnetic
phase diagram. For a magnetic field $H \parallel b$ axis the
magnetic phase diagram contains five different regions I (elliptical
helix) to V \cite{Willenberg2012,Schapers2013,Schapers2014} (see
Supplement). Region V displays very weak thermodynamic signatures,
and it was unclear, whether it is a distinct thermodynamic phase.

Here, we fully characterize its field-induced phases by means of
neutron diffraction (ND) and $^1$H nuclear magnetic resonance
(NMR). We establish the magnetic ordering vectors, and that region
V represents a thermodynamic phase. For phase V, we determine the
field dependence of the ICM SDW ordering vector and newly discover
complex states which might be understood in terms of phase
separation between MP and SDW$_p$ states.

ND was carried out using the single crystal instrument D10 at the
Institute Laue Langevin, France, and the instrument Wombat at ANSTO,
Australia. For the D10 experiment, the sample from a previous study
was used \cite{Willenberg2012}. A second single crystal of linarite
($9 \times 3 \times 0.5$\,mm$^3$) from the Grand Reef Mine, Arizona,
was used for the experiment on Wombat. The samples were placed in
cryomagnets with maximum field/base temperature of 6\,T/1.7\,K (D10)
and 12\,T/1.5\,K (Wombat), with the magnetic field applied along the
crystallographic $b$ axis. With this setup and a neutron wavelength
of 2.36\,\AA\ we were restricted along the $b$ direction to
$-0.25<k<0.25$ in reciprocal space for the D10 experiment, while for
the experiment on Wombat a wavelength of 4.61\,\AA\ was used
resulting in an access range $-0.19<k<0.19$.

$^1$H-NMR ($^1\gamma = 42.5749$\,MHz/T) studies were performed for
$T < 2.8$\,K using a phase-coherent Tecmag spectrometer in
combination with a He-flow cryostat. Frequency scans were
conducted down to 1.7\,K and at external fields $H \parallel b$
between 1.5\,T and 7.5\,T. The same single crystal was used as for
the D10 ND study. All NMR spectra were collected using a
$\pi/2-\tau-\pi$ Hahn spin-echo pulse sequence. The spectra have
not been corrected by the tiny spin-spin relaxation time $T_2 \sim
10 \mu$s.

\begin{figure}[b]
\centering
\includegraphics[width=0.8\columnwidth]{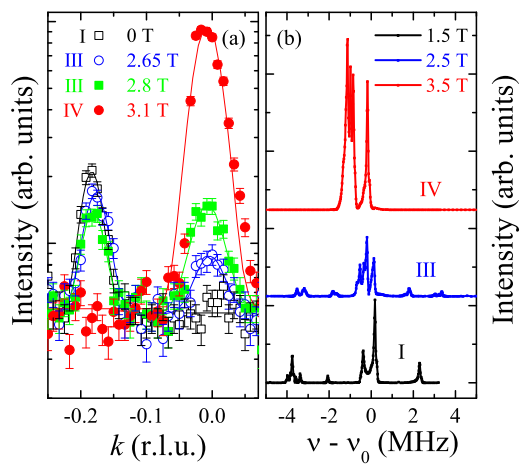}
\caption{a.) Neutron scattering scans for linarite along $k$ at
1.7\,K at different fields, reflecting the crossing from phase I
via III into IV. Solid lines are Gaussians fitted to the data to
determine peak positions. b.) NMR spectra for linarite at 1.7\,K
at different fields (spectra offset for clarity).}
\label{fig:fig1}
\end{figure}

By ND, scans of $(0~k~0.5)$ with varying $k$ were performed at
1.7\,K for fields up to 6\,T. For this temperature the sequence of
phases I--III--IV is traversed with increasing field. Consistent
with the 0\,T propagation vector, the magnetic Bragg peak
$(0~{-0.186}~0.5)$ is observed at low fields
(Fig.~\ref{fig:fig1}a). For increasing magnetic field, at the
boundary to phase III a second commensurate (CM) magnetic Bragg
peak appears at $(0~0~0.5)$, corresponding to spins coupled
parallel along the $a$ and $b$ axes and antiparallel along $c$. In
the field range 2.60 to 2.95\,T both peaks coexist, while for
higher fields (in phase IV) only the commensurate Bragg peak
remains.

The coexistence of two spin structures in phase III is also
observed in NMR. Fig.~\ref{fig:fig1}b displays the absolute
resonance change $\nu-\nu_0$ of the spectra taken in phase I, III,
and IV (for details see Supplement). The six NMR lines in phase I
are understood in terms of an ICM helical structure
\cite{Schapers2014}. In phase IV four discrete resonance lines
were observed. For linarite the only CM spin structure with four
resonance lines is an AFM, propagating along the $a$ or $c$
direction, consistent with ND. Conversely, in phase III in total
12 resonance lines are detected; four with high intensity at small
shifts $\nu - \nu_0 = -1.4$ to 0.5\,MHz, eight lines for shifts
larger than $\pm 1.5$\,MHz. The four central lines resemble those
of phase IV. The remaining eight peaks result from a modification
of the helical phase I structure. Thus, a phase separation occurs
into two spin structures. The two phases compete, as for
increasing field the intensity of the central lines increase while
the other decrease, reflecting a growth of the phase volume of the
first on behalf of the second. The same behavior is seen for the
field dependence of the corresponding Bragg peak intensities in ND
(Fig.~\ref{fig:fig1}a).

To determine the magnetic structure with propagation vector
$\boldsymbol{k}=(0~0~0.5)$ of phase IV, the intensity of 33
magnetic Bragg peaks (20 inequivalent) was measured at 4\,T by ND.
A refinement of the data ($R_F=9.5\,\%$) using the program
FullProf \cite{Rodriguez-Carvajal1993} reveals that the spins are
lying in the $ac$ plane, with an angle of $-27^{\circ}$ off the
$a$ axis (roughly parallel to $[10\bar{1}]$), the same as one of
the spin components of the phase I helix \cite{Willenberg2012}.
From the refinement an ordered moment of 0.79(1)$\mu_\text{B}$ per
Cu atom is derived. Similar refinements at 5.5\,T (20 inequivalent
Bragg peaks) yield the same spin structure with a moment of
0.73(2)$\mu_\text{B}$ per Cu atom ($R_F=12.5\,\%$). The decrease
of the AFM moment with field, and the observation of small
field-induced FM contributions on top of nuclear Bragg peaks,
reflects the development of field-induced spin polarization.

For the determination of spin structures in phase III, two sets of
magnetic Bragg peaks $(hkl)_M$ were collected at 2.8\,T using the
relation $(hkl)_M=(hkl)_N\pm \boldsymbol{k}$. For the ICM
structure the propagation vector $\boldsymbol{k} = (0~0.186~0.5)$
was used, for the CM structure $\boldsymbol{k} = (0~0~0.5)$. The
CM structure in phase III is refined using 15 peaks (14
inequivalent) with the phase IV spin model. The refinement of 18
inequivalent Bragg peaks of the ICM structure yields a circular
helix structure ($R_F=14.5$\,\%), where the moments of
0.64(2)$\mu_\text{B}$ lie roughly in the $bc$ plane.

In the related chain systems LiCuVO$_4$ and LiCu$_2$O$_2$,
applying magnetic field rotates the normal of the helical
structure parallel to the field. Here, such a spin flop of the
helix into the $ac$ plane is prohibited by the monoclinic angle
$\beta$. Instead the spins flop into a collinear spin structure in
the $ac$ plane. In phase III the spins start to flop into the $ac$
plane forming a collinear spin arrangement, while a coexisting
helical phase is retained. The fact that in phase III a circular
helix (spinning plane in the $bc$ plane) replaces the elliptical
helix reflects that for the latter it is energetically costly to
keep the large moment axis aligned along the field direction. For
larger fields all spins are flopped into the $ac$ plane forming
the collinear phase IV.

\begin{figure}[t]
\centering
\includegraphics[width=0.9\columnwidth]{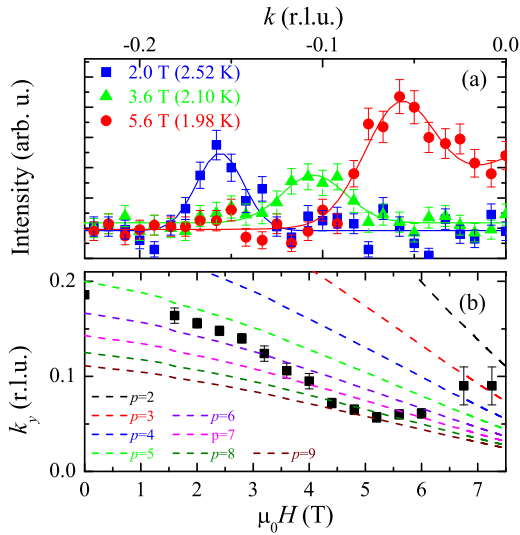}
\caption{a.) The magnetic Bragg peak position $(0~k~0.5)$ shifting
with field in phase V. b.) Field dependence of the propagation
vector in phase V compared to the 1D model SDW$_p$ states with $2
\leq p \leq 9$. Theory lines include the magnetization obtained from
$M(H,T)$ scans \cite{Schapers2013}.} \label{fig:fig2}
\end{figure}

Next, we have performed ND via $k$ scans through region V
(Fig.~\ref{fig:fig2}a). Surprisingly, and in spite of the very
weak signatures defining this region in thermodynamic measurements
\cite{Willenberg2012,Schapers2013}, we observe magnetic Bragg
peaks of the same width as the nuclear peaks. These peaks are even
observed in the intermediate field regime $\sim$\,4\,T, where in
thermodynamic measurements no anomalies were detected
\cite{Schapers2013}. It implies that region V is a distinct and
highly unusual thermodynamic phase. For magnetic structure
determination, a set of 8 inequivalent Bragg peaks was collected
at 6\,T. A refinement yields a sine-wave modulated structure, with
the spins aligned parallel to the $b$ axis ($R_F=7$\,\%).
Surprisingly, the SDW magnetic moment amplitude is only
0.44(1)$\mu_\text{B}$. This value is much smaller than what would
be expected from an extrapolation of the field dependence of the
magnetic moments measured in phase I and IV. Further, the
propagation vector $(0~{-k}~0.5)$ shifts in $k$ with field, as
shown in Fig.~\ref{fig:fig2}.

The sine-wave modulated structure with moments along the field
direction together with the shift of the $k$ value reminds of the
prediction of the longitudinal collinear SDW within hard-core
boson approximation \cite{Kecke2007,Sudan2009,Sato2013}, where the
shift depends on the number of bound magnons $p$ in the coexisting
or neighboring MP phase:

\begin{eqnarray}
\frac{k_y d}{\pi}= \frac{\left(1-M/M_\text{S} \right)}{p}.
\label{eq:1}
\end{eqnarray}

Here, $d$ denotes the distance of neighboring Cu spins along the $b$
axis, $M_\text{S}$ is the saturation magnetization
\cite{note1,note2}. To compare Eq.~\ref{eq:1} with the situation for
linarite, the curves with various $p$ values are included in
Fig.~\ref{fig:fig2}b. Surprisingly, at first glance no agreement is
found over a wide field range between the experimentally observed
evolution of $k_y$ and the theoretical prediction for a single chain
with fixed field independent exchange interactions for any fixed
value of $p$. However, taking into account the interchain coupling
we will arrive at a magnetization dependent reduced effective value
$\alpha_{\rm eff}(M/M_s)$. Then, the different $p$ values seen
experimentally can be qualitatively understood (see below and
Supplement).

\begin{figure}[b]
\centering
\includegraphics[width=0.95\columnwidth]{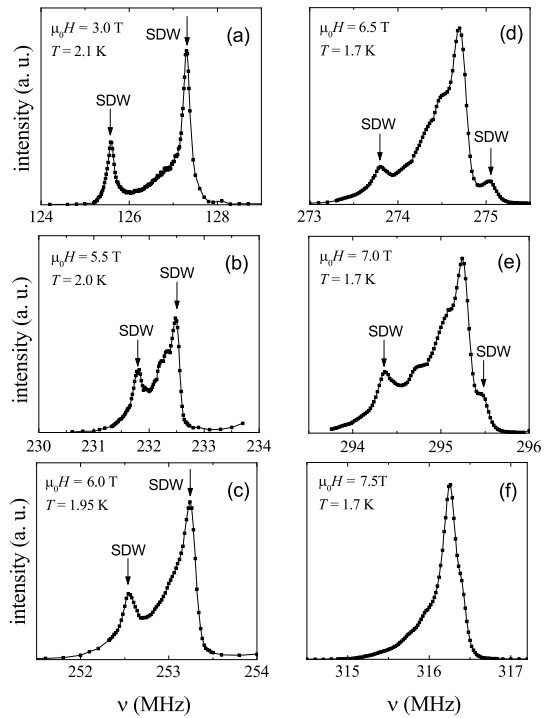}
\caption{The $^1$H-NMR spectra of linarite for different magnetic
fields and temperatures in phase V and in the paramagnetic
polarized state at 7.5\,T close to saturation.} \label{fig:fig3}
\end{figure}

Static magnetic order in phase V of linarite is also observed by
NMR. $^1$H-NMR frequency scans were performed in the field range
3\,--\,7.5\,T with an increment of 0.5\,T at different
temperatures. At 2.8\,K a paramagnetic signal is observed which is
composed of two almost overlapping lines from two inequivalent
$^1$H sites \cite{Schapers2014}. For fields $\leq 6$\,T, upon
lowering $T$ below a critical value $T_V$ the spectrum develops
horn-shaped NMR characteristics, that is two distinct peaks with a
finite intensity in between (Fig.~\ref{fig:fig3}a-c). It can be
accounted for by the SDW structure with only a magnetic component
along the $b$ direction (compare LiCuVO$_4$ \cite{Buttgen2014}).
The transition temperatures derived from NMR match those of phase
V obtained from thermodynamics for $H \parallel b$ axis
\cite{Willenberg2012,Schapers2013}, and define the phase boundary
in the field range 3.5\,--6\,T, where no transition has been
detected in thermodynamic quantities. Our findings imply that
phase V encloses all other magnetic phases (see Supplement).

Increasing the field to above 6\,T within phase V (in
Fig.~\ref{fig:fig3}d/e: 1.7\,K) produces a transfer of spectral
weight from the horn-shaped structure to a broadened two-peak
structure appearing in the middle of the SDW pattern. The shift of
the latter with increasing field follows the shift of the
paramagnetic polarized NMR signal close to saturation.
Qualitatively, this implies the presence of two different local
environments in the sample, {\it viz.}, a phase separation occurs.
In part of the sample there is SDW ordering producing the
horn-shaped NMR spectra. In contrast, the regions of the sample
exhibiting the broadened two-peak structure show no static magnetic
order. According to Refs.~\cite{Kecke2007,Hikihara2008,Sudan2009}
for coupled frustrated spin chains a field-induced transition from
the SDW$_p$ into a $p$-MP phase is expected in high fields. This
transition should appear as one from a magnetically long-range
ordered into one without static dipolar long-range order. Hence, we
suggest that the phase separation observed in phase V is related to
the transition from a SDW$_p$ phase into one with dominant MP
character in a quasi-1D material.

\begin{figure}[b]
\center
\includegraphics[width=1.0\columnwidth]{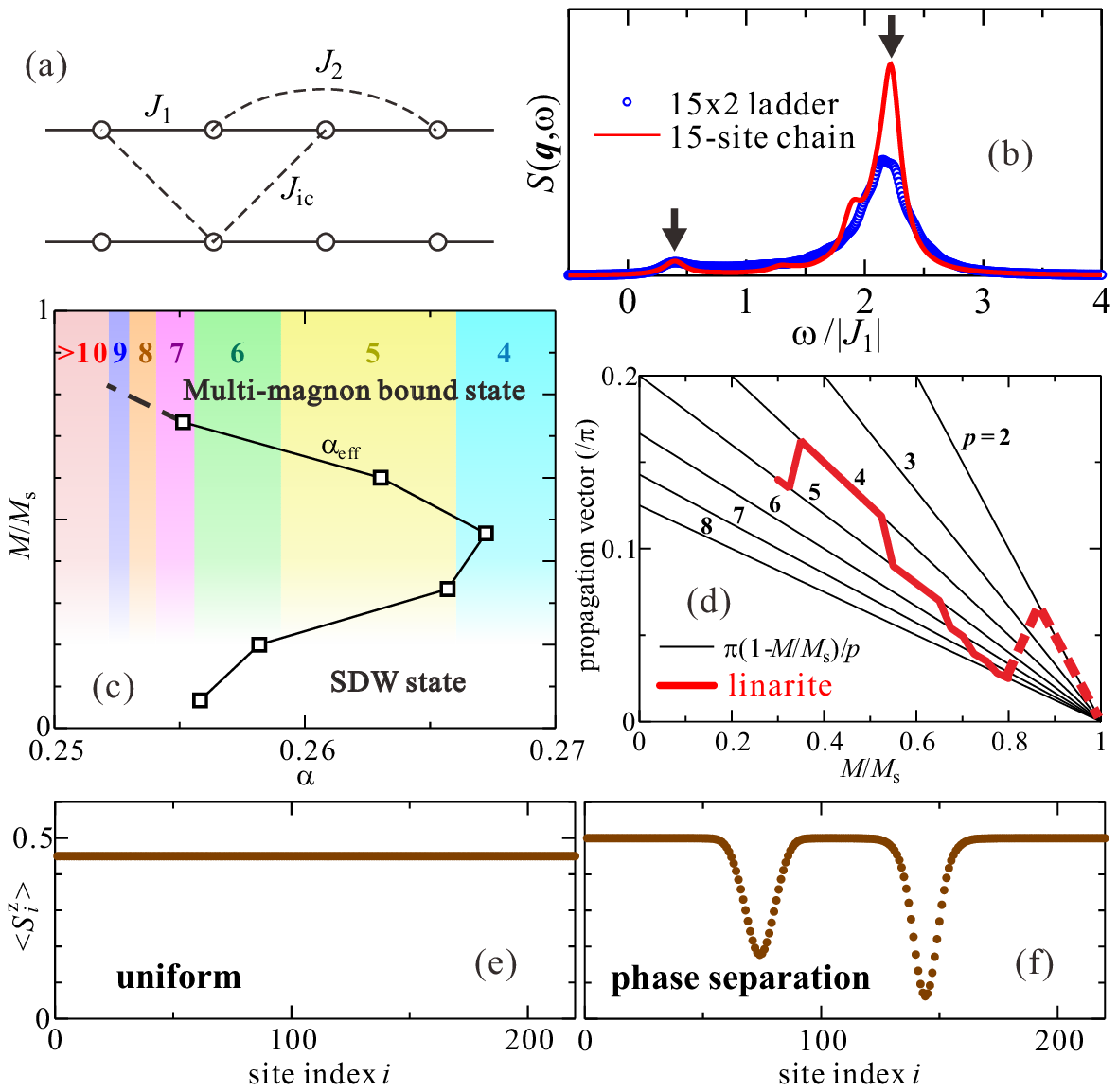}

\caption{(a) Lattice model of weakly-coupled $J_1 - J_2$ chains.
(b) Fitting of $S({\bf q},\omega)$ at $M/M_{\rm s} = 0.2$; between
${\bf q} = (\pi,\pi)$ for coupled two-chain with $\alpha = 0.36$,
$J_{\rm ic} = 0.1 |J_1|$ and $q = \pi$ for a single chain with
$\alpha_{\rm eff} = 0.257$. (c) Phase diagram of the multi-magnon
bound states and the estimated effective frustration ratio
$\alpha_{\rm eff}$ for linarite. (d) Suggested propagation vector
using Fig.~\ref{fig:fig4}(c). Local spin densities $\langle S^z_i
\rangle$ at $\alpha_{\rm eff} = 0.26$ and $M/M_{\rm s} = 0.95$ for
(e) periodic and (f) open chains.} \label{fig:fig4}
\end{figure}

To discuss our experimental results, we employ weakly-coupled $J_1
- J_2$ chains in a magnetic field $H$ along the $z$ axis. The
Hamiltonian reads as

\begin{eqnarray}
\nonumber
\hat{H} &=& J_1 \sum_{l,i} \mathbf{S}_{l,i} \cdot \mathbf{S}_{l,i+1} + J_2 \sum_{l,i} \mathbf{S}_{l,i} \cdot \mathbf{S}_{l,i+2} \\
&+& H \sum_{l,i} S_i^z + J_{\rm ic} \sum_{l,l^\prime,i,i^\prime}
\mathbf{S}_{l,i} \cdot \mathbf{S}_{l^\prime,i^\prime}, \label{hamJJ}
\end{eqnarray}
where $\mathbf{S}_{l,i}$ is a spin-$\frac{1}{2}$ operator at site
$i$ in chain $l$ and $J_{\rm ic}$ is a diagonal interchain
coupling (see Fig.~\ref{fig:fig4}(a)). As shown above, the ICM
propagation vector along the chain is $k_y = 0.186 \pi$ at $H =
0$; however, a single $J_1 - J_2$ chain with $\alpha = J_2/|J_1| =
0.36$ gives $k_y \approx 0.367 \pi$. This discrepancy can be
resolved by taking a specific diagonal $J_{\rm ic} \approx 10{\rm
K} = 0.1|J_1|$. Theoretically, the propagation vector in the
single $J_1$-$J_2$ chain is found from the maximum position of the
static spin-structure factor $S(k)$. Due to strong quantum
fluctuations and the resulting absence of static magnetic order in
1D with short range-couplings, only, it cannot be found from
$<S_i^z>$. But as a precursor, the former reflects nevertheless
the SDW modulations (induced by the coupling to neighboring chains
in 2D and 3D) we are looking for here. This maximum position of
$S(k)$ is reduced by decreasing $\alpha$ and it approaches 0 at
the FM critical point $\alpha_c= 1/4$. Such a reduction of the
propagation vector is also realized by increasing $J_{\rm ic}$ at
fixed $\alpha$ for a system of coupled chains (see Fig.~4a and
Fig.~5 in Ref.~\cite{Nishimoto2012}). It is thus interpreted that
the interchain coupling reduces the value of $\alpha$. Since a
single chain is more convenient than coupled chains for the
discussion of MP states, we first performed a mapping from two
coupled $J_1$-$J_2$ chains with $\alpha = 0.36$ and $J_{\rm ic} =
0.1 |J_1|$ with periodic perpendicular boundary conditions onto an
effective single $\tilde{J}_1$-$\tilde{J}_2$ chain with
$\alpha_{\rm eff} = -\tilde{J}_2/\tilde{J_1} = \alpha_{\rm
eff}(\alpha, M/M_s)$. For a wide range of the magnetization
$M/M_{\rm s}$ the values of $\alpha_{\rm eff}$ were estimated by
fitting the dynamical spin-structure factors $S({\bf q},\omega)$.
An example is shown in Fig.~\ref{fig:fig4}(b) (for more details
see Supplement). The estimated values of $\alpha_{\rm eff}$ are
plotted vs. $M/M_{\rm s}$ in Fig.~\ref{fig:fig4}(c). Note the
vicinity to $\alpha_c$.

Next, we found the number of bound magnons $p$ for a given
$\alpha$ by calculating the binding energy of a $p$-magnon bound
state near the saturation field, which is defined as

\begin{eqnarray}
\nonumber
E_{\rm b}(p)&=&\frac{1}{p}[E(S_z=S_{\rm max}-p)-E(S_z=S_{\rm max})]\\
&-&[E(S_z=S_{\rm max}-1)-E(S_z=S_{\rm max})], \label{binding}
\end{eqnarray}

where $E(S_z=S)$ is the ground-state energy with the $z$ component
of the total spin $S_z = S$, and $S_z = S_{\rm max}$ corresponds
to the fully polarized state. When the largest value of $E_{\rm
b}(p) (>0)$ is given by $p = p_{\rm max}$, we can prove that the
$p_{\rm max}$-magnon bound state is the most stable state;
whereas, if $E_{\rm b}(p)<0$ for all $p$ values no low-energy
magnon bound state exists. The results are shown in
Fig.~\ref{fig:fig4}(c). According to
Refs.~\cite{Kecke2007,Hikihara2008,Sudan2009}, the value of $p$
increases with approaching $\alpha_{\rm eff}=1/4$ and the region
of $\alpha_{\rm eff}$ becomes narrower for larger $p$. Based on
the relation between $p$ and $\alpha_{\rm eff}$, the propagation
vector of linarite is suggested to evolve similarly as shown in
Fig.~\ref{fig:fig4}(d) based on calculations within our effective
1D model, in semi-quantitative agreement with the experimental
data (Fig.~\ref{fig:fig2}(b)). For a brief discussion of the last
right dashed part of the broken red lines related to $p = 2$
obtained within an analogous $xyz$-anisotropic Heisenberg model
(to be published elsewhere) see Supplement.

The phase separation observed at high field may be attributed to
the low effective frustration ratio $\alpha_{\rm eff} \approx
1/4$, i.e., close to the critical point. This means that the FM
state is almost degenerate to other lower spin states. As an
illustration, the local spin densities $\langle S^z_i \rangle$
with periodic and open boundaries at $M/M_{\rm s} = 0.95$ for
$\alpha_{\rm eff} = 0.26$ are plotted in Figs.~\ref{fig:fig4}(e)
and (f), respectively. A uniform distribution is naturally
expected for periodic chains; whereas, interestingly, a phase
separation into partial polarized and unpolarized phases occurs by
open chain ends or possibly also near strong enough impurities in
the bulk being a common small disorder in real materials.

To conclude, linarite exhibits a field-induced behavior generic for
a FM-NN/AFM-NNN frustrated chain system. In addition, a two-step
spin-flop transition is present for external magnetic fields applied
along the $b$ axis. Further, a longitudinal sine-wave modulated
spin-structure phase encloses the other ordered phases. Here, at
relative low fields above the helical phase a shift in the
propagation vector qualitatively similar to LiVCuO$_4$ (with SDW$_2$
states, only) is observed. However, to the best of our knowledge we
report the first observation of several exotic (for Heisenberg
spin-1/2 systems) collinear longitudinal SDW$_p$ states (by changing
the external field), reaching even $p = 9$. We believe that this
result is related to the appropriate bare value of $\alpha$ and the
strong enough skew interchain coupling. Altogether, linarite appears
to be a good candidate to show MP behavior. A more detailed and
comprehensive study, its exotic SDW$_p$ states, and their interplay
with field-induced phase separation and exchange anisotropy provides
a challenge for future work. Also the comparison with other rare
cases of field-induced phase separation in frustrated zigzag chain
magnetic spiral systems as MnWO$_4$ \cite{Taniguchi2010} is of
interest for deeper insights in their complex physics including
multiferroicity.

Our work has been supported by the DFG under contracts WO 1532/3-1
and SU 229/9-1. We acknowledge fruitful discussions with W.\
Brenig, A.\ L\"auchli, U.\ R\"o\ss ler, N. Shannon, O.\ Starykh,
H. Tsunetsugu and M.\ Zhitomirsky. We thank G. Heide and M.
G\"{a}belein from the Geoscientific Collection in Freiberg for
providing the linarite crystal.

\bibliography{linarite_Neutronen+NMR}

\begin{thebibliography}{31}
\bibitem [{not({\natexlab{a}})}]{note1}%
  \BibitemOpen
  \href@noop {} {\bibfield  {journal} {\bibinfo  {journal} {Note that for
  linarite and LiVCuO$_4$ the lattice constant $b=2d$ due to alternating side
  groups.}\ } ({\natexlab{a}})}\BibitemShut {NoStop}%
\bibitem [{not({\natexlab{b}})}]{note2}%
  \BibitemOpen
  \href@noop {} {\bibfield  {journal} {\bibinfo  {journal} {The occurence of
  collinear SDW states with locally different collinear magnetization values is
  typical for intinerant magnetic systems but exotic for strongly correlated
  spin-1/2 systems where the magnitude of the local magnetization is fixed and
  the direction of the local moment changes spatially, i.e., leading either to
  non-collinear states or to various simple spin-up spin-down collinear
  Ne\'el-states.}\ } ({\natexlab{b}})}\BibitemShut {NoStop}%
\end{thebibliography}

\end{document}


\title{Supplementary Information for \\
Complex field-induced states in Linarite \lin with a variety of
high-order exotic SDW$_p$ states}
\author{B. Willenberg$^{1,2}$, M.\ Sch\"{a}pers$^{3}$, A.U.B.\ Wolter$^3$, S.-L.\ Drechsler$^3$, M.\ Reehuis$^2$, B.\ B\"{u}chner$^{3,4}$, A.J.\ Studer$^5$, K.C.\ Rule$^5$, B.\ Ouladdiaf$^6$, S.\ S\"{u}llow$^1$, and S.\ Nishimoto$^{3,4}$}

\address{
$^1$Institute for Condensed Matter Physics, TU Braunschweig, D-38106 Braunschweig, Germany\\
$^2$Helmholtz Center Berlin for Materials and Energy, D-14109 Berlin, Germany\\
$^3$Leibniz Institute for Solid State and Materials Research IFW Dresden, D-01171 Dresden, Germany\\
$^4$Institut f\"ur Festk\"orperphysik, TU Dresden, D-01062 Dresden, Germany\\
$^5$The Bragg Institute, ANSTO, Kirrawee DC NSW 2234, Australia\\
$^6$Institute Laue-Langevin, F-38042 Grenoble Cedex, France}

\maketitle
\date{\today}

In this Supplementary part we present the magnetic phase diagram of
linarite \lin with $H \parallel b$ axis for the convenience of the
reader once again, but now extended by the inclusion of new NMR data
and under consideration of additional neutron scattering data taken
in magnetic fields. These new data establish phase V as a distinct
thermodynamic phase. Furthermore, the evolution of the $^1$H-NMR
spectra in phase III is shown as a function of the magnetic field.
The nature of the different subspectra are discussed and compared to
the $^1$H-NMR spectra in the phases I and IV. Finally, we explain in
more detail the mapping procedure of two coupled $J_1 - J_2$,
$J_{ic}$-chains onto an effective single $\tilde{J}_1$-$\tilde{J}_2$
chain but with a magnetization dependent frustration ratio
$\alpha_{\rm eff} = \alpha_{\rm eff} \left( \alpha \left[ M/M_s
\right] \right)$.

\section{The magnetic phase diagram of linarite for $H \parallel b$}

The phase diagram for a magnetic field $H$ applied parallel to the
crystallographic $b$ axis of linarite contains five different
regions/phases I to V
\cite{Willenberg2012,Schapers2013,Schapers2014} (see Fig.~S1). The
ground state phase I below $\sim 2.5$\,T is formed by an
incommensurate (ICM) elliptical helix (ordering wave vector
$\boldsymbol{k}=(0~0.186~0.5)$, moment size $\mu_{ord} \sim 0.64$ to
0.83\,$\mu_B$). The field-induced transition from phase I into phase
IV involves either a hysteretic transition (region II, below $\sim
0.6$\,K), a direct one (up to $\sim 1.25$\,K), or the transit
through a phase III (up to $\sim 1.9$\,K and 3.2\,T). Region V (up
9.5\,T) displays only very weak thermodynamic signatures, therefore
it was not clear whether it is a distinct thermodynamic phase. This
important issue has been settled by $^1$H-NMR and neutron
diffraction measurements presented in this manuscript. The
transition temperatures into phase V obtained from the temperature
dependence of our $^1$H-NMR spectra are included in Fig.~S1 for the
first time.

\begin{figure}[h]
\center
\includegraphics[width=0.55\columnwidth]{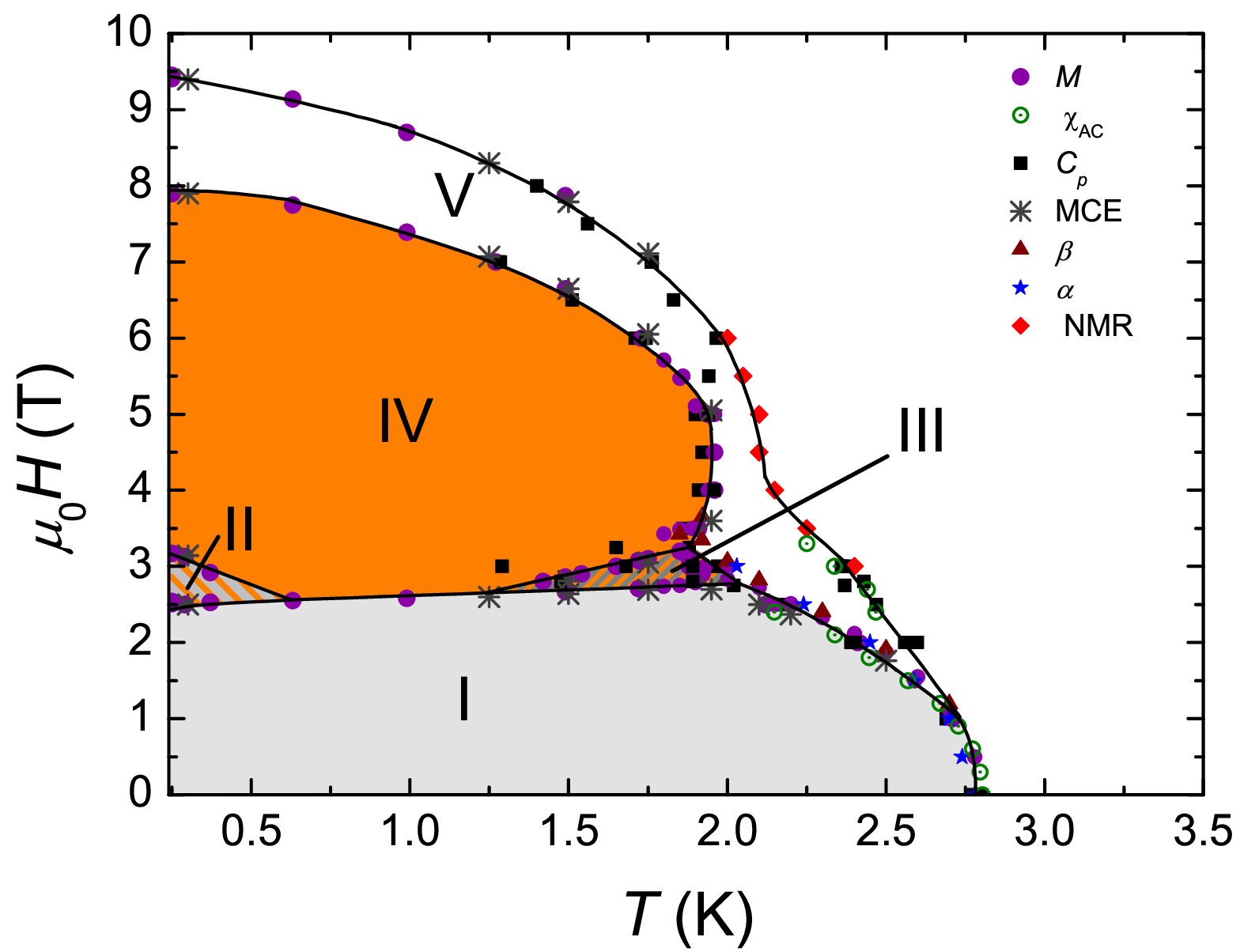}
\caption{Phase diagram of linarite with the applied field along the
crystallographic $b$ axis. For a brief description of the phases I
to V see text and Refs.~1-3. Transition temperatures obtained from
new $^1$H-NMR data are depicted as red diamonds.}
\label{fig:suppl_Fig3}
\end{figure}

\section{Nature of the $^1$H-NMR spectra in phase III}

Phase III is wedged in between phase I and phase IV in the
magnetic phase diagram of linarite for $H \parallel b$. From the
thermodynamic studies it was speculated that phase III is part of
a two-step spin spiral reorientation process by going from phase I
via phase III to phase IV. This notion needs to be verified on a
microscopic level by means of a combined NMR and neutron
scattering study. Here, additional information on the NMR part is
presented.

\begin{figure}[b]
\center
\includegraphics[width=0.7\columnwidth]{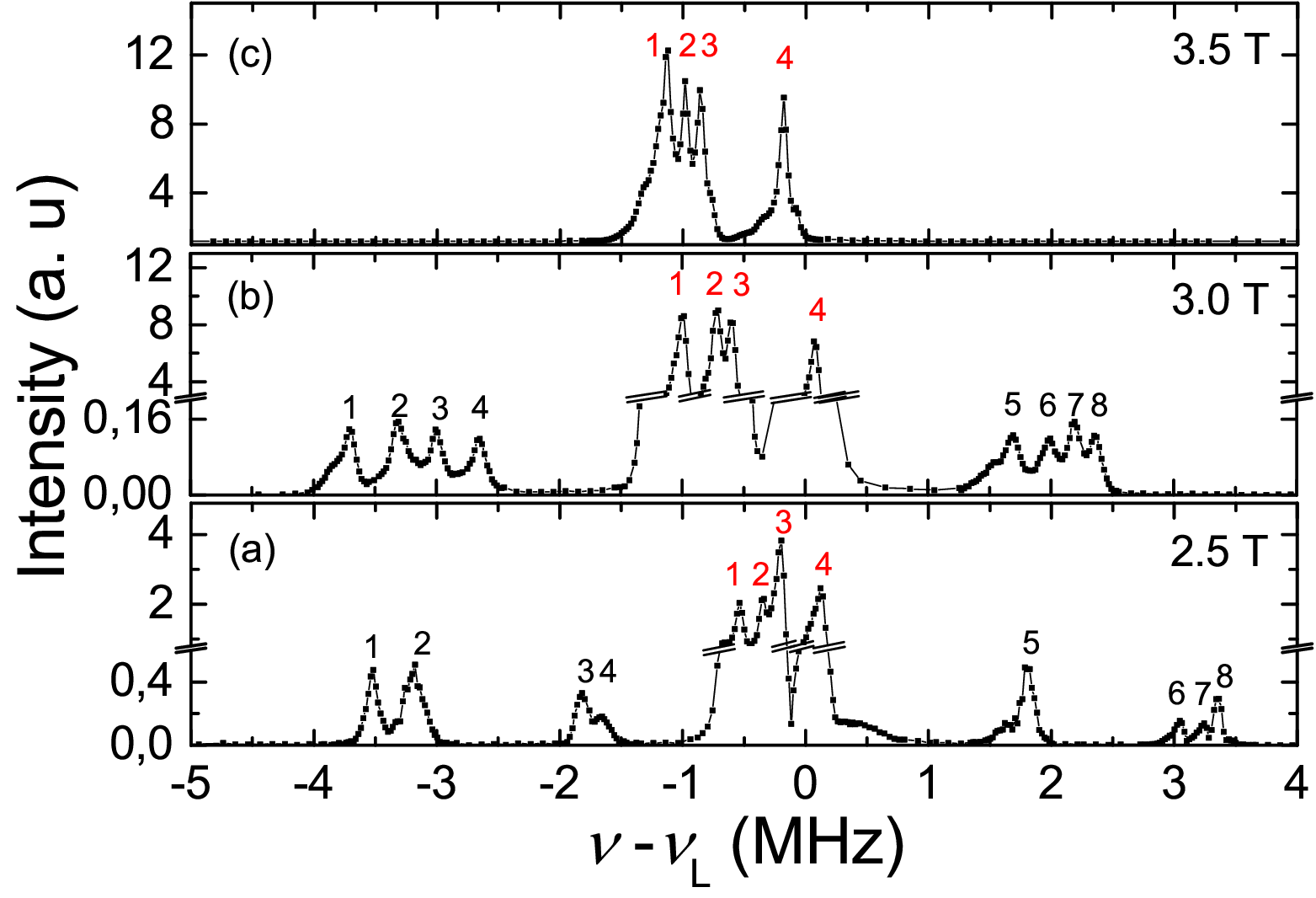}
\caption{(a) and (b) $^1$H-NMR spectra of linarite in phase III at
$T = 1.7$\,K for $\mu_0H$ = 2.5 and 3.0\,T $\parallel b$ axis: Four
discrete lines with high intensities (arbitrarily labeled by red
numbers) are visible in the middle of each spectrum, which can be
associated to the spectra known from phase IV. Eight additional
resonance lines (arbitrarily numbered in black) are placed at the
border of each spectrum, belonging to a new magnetic structure but
which is closely related to the helical magnetic ground state. (c)
For comparison, $^1$H-NMR spectrum of linarite in phase IV at $T =
1.7$\,K for $\mu_0H$ = 3.5\,T $\parallel b$ axis.}
\label{fig:suppl_Fig2}
\end{figure}

$^1$H-NMR was measured at 1.7\,K in 2.5 and 3.0\,T and is shown in
Fig.~\ref{fig:suppl_Fig2}(a) and (b). For both fields 12 resonance
lines are detected, four of them with a high intensity in the
middle of the spectra close to the absolute resonance change $\nu
- \nu_\text{L} = 0$\,MHz, and eight lines situated at the borders
of the spectra. The four high-intensity lines in the middle of the
spectra closely resemble the shape of the spectra in phase IV,
with the latter shown in Fig.~\ref{fig:suppl_Fig2}(c) for an
external field of 3.5\,T. In turn, by increasing the external
magnetic field from 2.5 to 3\,T, the whole spectrum shifts to
lower frequencies $\nu - \nu_\text{L}$.

Remarkably, the four lines in the middle gain intensity with
increasing field, whereas the remaining eight peaks lose
intensity. At 2.5\,T the four peaks have a relative intensity of
76\,\%, which is increased to 96\,\% at 3.0\,T, as determined by
integrating the subspectra and assuming constant $T_2$ times in
2.5 and 3\,T for the individual peaks.

The eight-peak subspectrum appears to be discrete, which would point
towards an underlying commensurate magnetic structure. However, also
here the spin-spin correlation time $T_2$ is very short, which would
hamper identification of a horn-shape spectrum expected for an
incommensurate spiral structure (cf. the situation for phase I
\cite{Schapers2014}). Our neutron scattering data in phase III
provide evidence for a coexistence of the phase IV commensurate
magnetic structure with an incommensurate spin spiral. Then, on the
supposition of horn-shaped spectra with pairs of lines reaching from
one to the other border of the spectrum, viz., from negative to
positive $\nu - \nu_\text{L}$, the intensity in between the peaks is
easily suppressed due to $T_2$ being of the order of 10\,$\mu$s,
resulting in the appearance of seemingly discrete peaks.

Under the assumption that the eight remaining peaks are related to
the incommensurate helical ground state (due to the same number of
resonance lines), a modification of this structure may be expected
at the phase boundary I--III. The abrupt change observed in $\nu -
\nu_\text{L}$ at the phase boundary from phase I to III makes a
change of the spin structure mandatory. Modifications could be a
significant change in (either) the spinning plane, the magnetic
moment, or the propagation vector. Hence, the NMR data is fully
consistent with the conclusion from neutron scattering, i.e., a
coexistence in the sample of a spin spiral with phase IV regions.

\section{Mapping of two coupled $J_1 - J_2$, $J_{\rm ic}$ chains onto an effective single $\tilde{J}_1 - \tilde{J}_2$ chain}

\begin{figure}[b]
\center
\includegraphics[width=0.7\columnwidth]{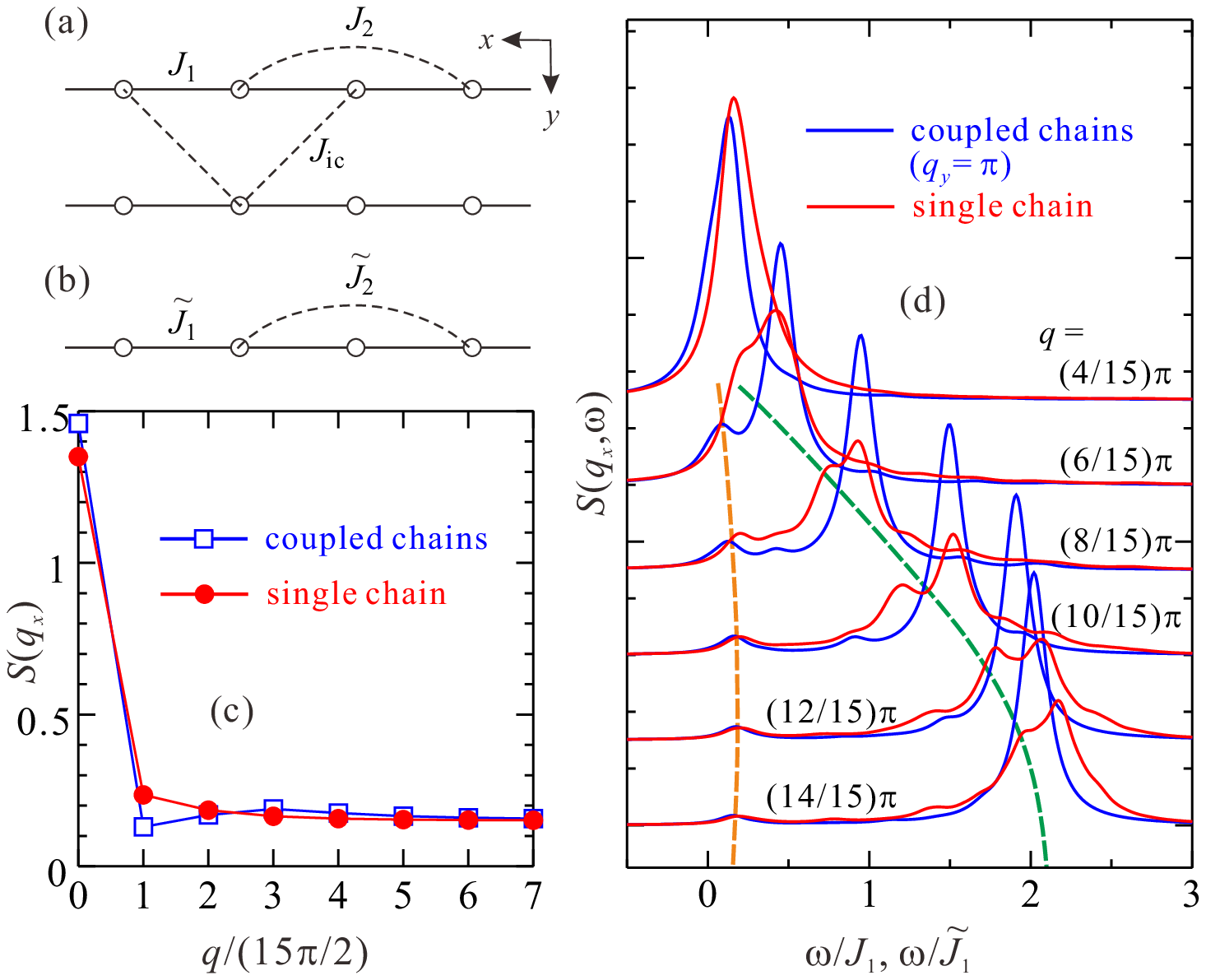}
\caption{Lattice models of (a) weakly-coupled $J_1 - J_2$ chains
and (b) effective $\tilde{J}_1 - \tilde{J}_2$ chain. (c), (d)
Comparison of the static and dynamical spin structure factors
between the weakly-coupled $J_1 - J_2$ chains with $\alpha =
J_2/|J_1| = 0.36$, $J_{\rm ic}/|J_1| = 0.1$ (blue) and the
effective single $\tilde{J}_1 - \tilde{J}_2$ chain with
$\alpha_{\rm eff} = \tilde{J}_2/|\tilde{J}_1| = 0.263$ (red) for a
fixed magnetization $M/M_{\rm s} = 0.6$. The static structure
factor of the coupled chains $S(q_{\rm x})$ has been averaged over
$q_{\rm y} = 0$ and $q_{\rm y} = \pi$. The orange and green dotted
lines denote the lower and upper bounds of the spinon continuum,
respectively.} \label{fig:suppl_fig1}
\end{figure}

In the main text, we have performed a mapping from the $J_1 - J_2$
chains weakly coupled by the diagonal interchain exchange
interaction $J_{\rm ic}$ (see Fig.~\ref{fig:suppl_fig1}(a)),
\begin{eqnarray}
\hat{H} = J_1 \sum_{l,i} \mathbf{S}_{l,i} \cdot \mathbf{S}_{l,i+1}
+ J_2 \sum_{l,i} \mathbf{S}_{l,i} \cdot \mathbf{S}_{l,i+2} + H
\sum_{l,i} S_{l,i}^z + J_{\rm ic} \sum_{l,l^\prime,i,i^\prime}
\mathbf{S}_{l,i} \cdot \mathbf{S}_{l^\prime,i^\prime},
\label{hamJJ}
\end{eqnarray}
onto an effective single $\tilde{J}_1 - \tilde{J}_2$ chain (see
Fig.~\ref{fig:suppl_fig1}(b)),
\begin{eqnarray}
\hat{H} = \tilde{J}_1 \sum_i \mathbf{S}_i \cdot \mathbf{S}_{i+1} +
\tilde{J}_2 \sum_i \mathbf{S}_i \cdot \mathbf{S}_{i+2} + H \sum_i
S_i^z, \label{hamJsingle}
\end{eqnarray}
where $\mathbf{S}_{l,i}$ is a spin-$\frac{1}{2}$ operator at site
$i$ in chain $l$. The objective of this mapping is to mimic the
internal magnetic properties of a chain in the coupled $J_1 - J_2$
model by an effective single $\tilde{J}_1 - \tilde{J}_2$ chain. A
reliable mapping can be achieved by calculating the dynamical spin
structure factor
\begin{eqnarray}
S(\mathbf{q},\omega) = \sum_{\nu>0} |\langle
\psi_\nu|S(\mathbf{q})|\psi_0\rangle|\delta(\omega-(E_\nu-E_0)),
\label{Sqw}
\end{eqnarray}
where the static structure factor $S(\mathbf{q})$ is simply the
Fourier transform of $S_{l,i}^z$, $E_\nu$ and $|\psi_\nu \rangle$
are the $\nu$-th eigenenergy and eigenstate of the system,
respectively (the ground state denoted by $\nu = 0$). For some
fixed magnetizations $M/M_{\rm s}$ we determined the value of
$\alpha_{\rm eff} = \tilde{J}_2/|\tilde{J}_1|$ to reproduce the
spin structure factor of the $15 \times 2$ coupled chains by the
effective $15$-site chain as much as possible. As an example of
the mapping, the comparison of the static and dynamical structure
factors between the original coupled chains and the effective
single chain at $M/M_{\rm s} = 0.6$ is shown in
Fig.~\ref{fig:suppl_fig1}(c) and (d). We can see a good agreement.

Notice that the obtained "renormalized" $\alpha_{\rm eff}$ depends
naturally on both the bare value $\alpha$ and on the type and
strength of the interchain coupling. The smaller the former and the
larger the skew interchain couplings are, the smaller is the
resulting $\alpha_{\rm eff}$. To illustrate this point, we mention
that for larger bare $\alpha$ values above 0.7, as relevant for
LiVCuO$_4$, one remains in the $p = 2$-region for all magnetic
fields in accord with the data of Mourigal {\it et al.}
\cite{Mourigal2012}. In contrast, for Li$_2$CuO$_2$ with a smaller
bare $\alpha \approx 0.32$ as compared to linarite one reaches
already the ferromagnetic phase with $\alpha_{\rm eff} < 0.25$
\cite{Lorenz2009} in accord with the observed alignment of magnetic
momenta along the chains and without any anomalous longitudinal
collinear SDW$_p$ states. Finally, since the pitch is only weakly
affected by the exchange anisotropy, the inclusion of the last
"zigzag" into the curve shown in Fig.~4(d) of the main text obtained
from an anisotropic generalization of our single chain Hamiltonian
is reasonable (see also the next subsection of the present
Supplement).

\section{Nematic ($p=2$) state near the saturation}

According to preliminary (unpublished) measurements for our title
compound the material is highly anisotropic (see also
Ref.~\cite{Wolter2012}) and a more general spin model might be
provided by a Heisenberg chain with $xyz$-exchange anisotropy for
any $\alpha_{\rm eff}$. Then the Hamiltonian is written as
\begin{eqnarray}
H = \sum_{i,\gamma=x,y,z} \tilde{J}_1^\gamma
S^\gamma_iS^\gamma_{i+1} + \tilde{J}_2 \sum_i \mathbf{S}_i \cdot
\mathbf{S}_{i+2} + H \sum_i S_i^z, \label{hamxyz}
\end{eqnarray}
where $S^\gamma_i$ is the $\gamma$ component of $\mathbf{S}_i$. The
$xy$ components of the first term can be divided into an exchange
term $\frac{\tilde{J}_1^x + \tilde{J}_1^y}{4}(S^+_iS^-_{i+1}+h.c.)$
and a double spin-flip term $\frac{\tilde{J}_1^x -
\tilde{J}_1^y}{4}(S^+_iS^+_{i+1} + h.c.)$. This term has the same
form as the order parameter of the nematic state $\langle
S^-_iS^-_{i+1} \rangle$. It might be dominant at high magnetization
$M/M_{\rm s} \sim 1$. Therefore, qualitatively a nematic state might
be expected near the saturation in the presence of an $xyz$-exchange
anisotropy. More details will be published elsewhere.

\bibliography{linarite_Neutronen+NMR}